\documentstyle[preprint,aps,prl]{revtex}
\tightenlines

\begin{document}
\draft
\preprint{zhou.tex}
\title{Bending and Twisting Elasticity:$\;\;$ a Revised  Marko-Siggia
Model on DNA Chirality}

\author{Zhou    Haijun$^{1}$ and   Ou-Yang      Zhong-can$^{1,2}$}
\address{
$^{1}$Institute of Theoretical Physics, Chinese Academy of Sciences,\\
P.O. Box 2735, Beijing 100080, China\\
$^{2}$Center for Advanced Study, Tsinghua University, Beijing 100084, China}
\date{\today}

\maketitle
\begin{abstract}
 A revised Marko-Siggia  elastic model for DNA double helix 
 [Macromolecules {\bf 27}, 981 (1994)] is proposed, which includes 
 the WLC bending energy and a new chiral twisting energy 
 term.  It is  predicted that 
 the mean helical repeat length (HRL) for short DNA rings
 increases with the decreasing of chain 
length; while for
very long chains, their mean HRL is the same, independent of both the
chain length and   whether the ends are  closed, it
 is longer than the value for rectilinear DNAs. Our results are
 in good agreement with experiments.
 \end{abstract}

\narrowtext

Single-molecule extension experiments on DNA molecules show that freely
fluctuating open chains (FFOCs) could be well described by the 
inextensible worm-like chain (WLC) model [1-3]. 
For a WLC chain of total length $L$, its intrinsic elastic energy is of the
form $\beta E=\int_{0}^{L} A \kappa^{2} ds /2$, where $A \simeq 150$ base pair
(bp)
 is called the bending persistence length and $\kappa= |\partial_s {\bf t}|$ is 
the curvature, the change rate of the tangent unit vector
${\bf t}(s)$ at arc-length $s$, $\beta=1/k_{B}T$ with $k_{B}$ being the
Boltzmann constant and $T$  the environment temperature [3]. However, in addition
to bending ones, double stranded DNA molecules have also
twisting degrees of freedom, and  the total intrinsic energy formula
 for a deformed DNA chain is still unclear [4-10]. To know the
exact form of  the energy formula is very important for the
study of DNA configurational properties,
especially in the case of torsionally constrained DNAs, such
as covalently closed DNA rings in cells.  For example, it has been
widely accepted that bending elasticity and twisting elasticity determine to
a large extent the particular tertiary structures of DNA rings [4-6].
Previous studies often regard a DNA chain as a thin  elastic rod with 
isotropic cross section [11], the total elastic energy is assumed to be
\begin{equation}
\beta E_a=\int_{0}^{L} [ {A\over 2} (\Omega_1^2+\Omega_2^2)+{C\over 2}
 (\Omega_3-\omega_0)^{2} ] ds
 \end{equation}
  with bending and twisting deformations being independent of each other,
 here $C$ is called the twisting persistence length and $\omega_0$ is 
the spatial angular frequency of the unstressed
DNA double helix [4-6], and $\Omega_1^2+\Omega_2^2=\kappa^2$ [11].
 Although this simple achiral model is useful in some cases 
and it seems to be the most natural extension of  
  the already verified WLC model, it can not
properly describe the chiral characteristics of 
real DNA chains, this chirality of DNA molecules was  clearly demonstrated
by  the single-molecule experiment of Strick {\it et al}. [7].
Another very  important 
phenomenon related to the chirality of DNA molecules is that
Nature prefers $^{``}$linking number deficit$^{"}$ in circular DNAs [12-14],
this bias can not be well explained by the achiral model  (1) 
studied previously [4-6].

Recently, a chiral elastic theory was proposed by Marko
 and Siggia to incorporate
coupling between bending and twisting deformations in the energy formula [8].
 This 
model was based on a careful consideration of the intrinsic symmetry of DNA
chains. Later on, Kamien {\it et al}. extended the Marko-Siggia (MS) model to
investigate twist$-$stretch coupling of highly extended DNA supercoils and
found good agreement with experiment [9, 10]. 
However, when applying the MS model to the case of FFOCs, we find that
it is in general not compatible with the already verified WLC theory. 
To overcome this shortcoming, in our present work we first propose a revised
version of the MS model to ensure this compatibility. It is shown that this 
can be attained if we further hypothesize that unstressed circular DNAs
form flat circles. The corresponding internal deformation energy
is also consisted of two parts as in Eq.~(1), the bending energy and
the twisting energy.  The only difference is that $\Omega_3$ in (1) is
replaced by $\Omega_3+(B/C)\Omega_1$ in the twisting energy.
After  deriving the new elastic energy, we use this model to discuss the
mean helical repeat length (HRL) of open and closed DNA chains. 
Our result show that for short ring-shaped DNAs, the
shorter the chain, the longer its HRL. This tendency is consistent with 
experimental observations. For very long chains with twisting freedom, i.~e.,
open chains or  closed chains with at least one  defect,  their
mean HRL is independent of both the chain length and whether the ends are closed
or not, its value is longer than that of  rectilinear DNAs.

Experimentally, it was  discovered that random solution DNAs have
a mean HRL significantly longer that of rectilinear DNA fibers [15, 16].
 However,
the reason for this discrepancy has been obscure for quite a long time. Some 
researchers suggested that maybe ionic concentrations differ in fibriform and
solution DNAs, causing a observable effect on the HRL. 
Our present theoretical results indicate another possibility: that 
the chirality  of DNA might be the real reason for this  discrepancy.
In fact, we find that the ionic conditions in DNA fibers with high humidity
and in solution DNAs differ only slightly in the experiment of Refs.~[16] and
[15], so we feel our present explanation may be more reasonable. 

First we briefly review the main points of the MS model [8-10].
 The configurations of an inextensible polymer are specified by three
orthonormal unit vectors $\{{\bf u}(s), 
{\bf n}(s),{\bf t}(s) \}$ along the chain,
where $\bf{t}$ is the axial direction vector of the DNA double helix and
$\bf{u}$  is a unit vector perpendicular to $\bf{t}$ and pointing from
one back-bone chain to the other, $\bf{n}=\bf{t}\times\bf{u}$.
 It proves to be convenient to use Euler angles by setting 
 ${\bf u}={\bf e}_1$, ${\bf n}={\bf e}_2 $
 and ${\bf t}={\bf e}_3$, with $\partial_s {\bf e}_i 
 ={\bf \Omega}\times {\bf e}_i$ 
($i=1, 2, 3$), here ${\bf \Omega}=(\Omega_1, \Omega_2, \Omega_3)$ is angular
velocity of the frame $\{{\bf e}_i\}$ [11].
 Symmetry analysis shows that the polymer's
 properties should remain unchanged under the transformation 
$\{{\bf e}_1\rightarrow -{\bf e}_1$, ${\bf e}_3\rightarrow - {\bf e}_3\}$,
 thus, under the fundamental assumption
that {\it an undistorted open DNA is a linear double helix with
spatial frequency $\omega_0$}, the most
general elastic energy up to quadratic
order in the deformations should be of the following form [8]
\begin{equation}
\beta E_c=\int ds [{\frac{A^{\prime}}{2}}\Omega_{1}^{2}+{\frac{A}{2}}\Omega_{2}^{2}+{\frac{C}{2}}
(\Omega_3-\omega_0)^{2}+B \Omega_1 (\Omega_3-\omega_0)]
\end{equation}
the first two terms are related to bending deformations and $A^{\prime}$, $A$ 
are bending persistence lengths along the 
directions $\bf{e}_2$ and $\bf{e}_1$, respectively, 
the third term is twisting energy, the last
term is caused by bend$-$twist coupling and $B$ is the coupling constant [8].
Eq.~(2) can be further extended to include  
stretch$-$twist coupling [9, 10], but this 
effect is not important for DNAs at average conditions.

 It is easy to know that, for FFOCs who can  twist freely so we need only to
 consider bending deformations, the MS model (2) 
 is not equivalent with the WLC internal energy
$\int A(\Omega_1^{2}+\Omega_{2}^{2}) ds/2$,  unless the condition
 $B^{2}= (A^{\prime}-A)C$ is satisfied perfectly [see Eq.~(10)].
 In other words, if we assume the 
correctness of model (2)  and regard it as the  starting point,
it will not lead to the WLC energy for freely-twisting chains, unless
$B^{2}=(A^{\prime}-A)C$.  However, the WLC model for open DNA chains  has
been confirmed by experiments [1-3], so for the MS model to be appropriate the 
bend$-$twist coupling constant should satisfy this strict requirement.
However, one would wonder why should  the twist$-$bend coupling $B$ be 
completely determined by bending and twisting characteristic lengths?
Is there any intrinsic reason? In the following we 
 will see this relation is actually implied in the
MS model, but we need to add in the MS theory another fundamental
assumption that {\it  an undistorted  closed
DNA ring should form a flat circle}.

To show this, let's consider the case of short chains. 
For a short DNA chain (less than about one bending persistence length),  entropic elasticity caused by thermal fluctuations is neglectable
and the chain's free energy equals its internal deformation energy Eq.~(2). 
The first variation $\delta^{(1)}\{\beta E_c+{\bf{\lambda}}\cdot 
\int_{0}^{L} {\bf t}ds \}=0$ with respect to the three Euler angles 
$\theta$, $\phi$ and $\psi$ [11] gives the equilibrium shape equation, here $
{\bf{\lambda}}=(\lambda_1, \lambda_2, \lambda_3)$ is a Lagrange multiplier to
take account the possible constraint of fixed end-to-end distance. 
 The shape equation
is composed of the following three equations [17]
\begin{eqnarray}
&&A^{\prime}( \phi^{\prime}\sin\psi \sin\theta+\theta^{\prime}\cos\psi)
\phi^{\prime} \sin\psi \cos\theta-A^{\prime} [ (\phi^{\prime} \sin\psi \sin\theta+
\theta^{\prime} \cos\psi) \cos \psi]^{\prime} \nonumber \\
&&+A(\phi^{\prime} \cos\psi \sin\theta-\theta^{\prime}\sin\psi)\phi^{\prime}\cos\psi\cos\theta +A [(\phi^{\prime} \cos\psi \sin\theta - \theta^{\prime} \sin\psi)
\sin\psi]^{\prime} \nonumber \\
&&-C (\phi^{\prime}\cos\theta+\psi^{\prime}-\omega_0)\phi^{\prime}\sin\theta+
B(\phi^{\prime} \cos\theta+\psi^{\prime}-\omega_0)\phi^{\prime}\sin\psi \cos\theta \nonumber \\
&&-B[(\phi^{\prime} \cos\theta+\psi^{\prime}-\omega_0) \cos\psi]^{\prime}-
B (\phi^{\prime} \sin\psi \sin \theta+\theta^{\prime} \cos\psi) \phi^{\prime}\sin\theta+\lambda_1 \sin\phi \cos\theta \nonumber \\
&&-\lambda_2 \cos\phi \cos\theta -\lambda_3 \sin\theta =0, \\
&& \; \nonumber \\
&&-A^{\prime} [(\phi^{\prime} \sin\psi \sin\theta+\theta^{\prime} \cos \psi) \sin\psi \sin\theta]^{\prime}-
A [(\phi^{\prime}\cos\psi \sin\theta-\theta^{\prime} \sin\psi)
\cos\psi \sin\theta]^{\prime} \nonumber \\
&&-C[(\phi^{\prime}\cos\theta+\psi^{\prime}-\omega_0)
\cos\theta]^{\prime}-B [(\phi^{\prime} \cos\theta+\psi^{\prime}-\omega_0)\sin\psi \sin\theta]^{\prime} \nonumber \\
&&-B[(\phi^{\prime}\sin\psi \sin\theta+\theta^{\prime}\cos\psi)\cos\theta]^{\prime}+\lambda_1 \sin\theta\cos\phi+\lambda_2 \sin\theta \sin\phi =0, \\
&& \; \nonumber \\
&&A^{\prime} (\phi^{\prime}\sin\psi \sin\theta+\theta^{\prime}\cos\psi)(\phi^{\prime}\cos\psi \sin\theta-\theta^{\prime} \sin\psi) \nonumber \\
&&+A (\phi^{\prime}\cos \psi\sin\theta-\theta^{\prime}\sin\psi)(-\phi^{\prime}\sin\psi\sin\theta-\theta^{\prime}\cos\psi) \nonumber \\
&&-C(\phi^{\prime}\cos\theta+\psi^{\prime}-\omega_0)^{\prime}-B(\phi^{\prime}\sin\psi\sin\theta+\theta^{\prime}\cos\psi)^{\prime} \nonumber \\
&&+B (\phi^{\prime}\cos\theta+\psi^{\prime}-\omega_0)(\phi^{\prime}\cos\psi\sin\theta-\theta^{\prime}\sin\psi) =0.
\end{eqnarray}
In Eqs.~(3-5),
 differentiation with respect to arc length $s$ are denoted by superscript
prime  (it should not be confused with that of the elastic constant
 $A^{\prime}$).
Eqs.~(3-5) determine bending and twisting manners of short DNA chains in 
equilibrium. To check the validity of these equations,
we see that (3-5) require that the linear equilibrium shape
 with $\theta^{\prime}=\phi^{\prime}
=0$ should have $\psi^{\prime}=\omega_0$,  and   its HRL is $h_0=2\pi/\omega_0$.
This is in agreement with the previously mentioned fundamental assumption of the
MS model [8], so Eqs.~(3-5) are valid.

Now we investigate whether a flat-circular configuration can be a equilibrium one.
 For a flat circle (its two Euler angles can
 be set to be   $\theta=\pi/2$ and
$\phi^{\prime}=p=2\pi/L$ for convenience) to be a equilibrium shape, 
Eqs.~(3-5) 
require that the third angle, $\psi$, should satisfy the following  equations,
\begin{eqnarray}
&&(A-A^{\prime}) p (\sin\psi \cos\psi)^{\prime}-C p (\psi^{\prime}-\omega_0)-
B[(\psi^{\prime}-\omega_0)\cos\psi]^{\prime}-B p^{2}\sin\psi-\lambda_3=0, \\
&&(A^{\prime}-A) p^{2} \sin\psi \cos\psi-C \psi^{\prime\prime}-B \omega_0
p \cos\psi =0,\\
&&2 (A-A^{\prime} )p\psi^{\prime}\sin\psi\cos\psi-B(\psi^{\prime})^{2}\cos\psi+
B\omega_0 \psi^{\prime}\cos\psi-B\psi^{\prime\prime}\sin\psi \nonumber \\
&&\;\;\;\;\;\;\;\;\;\;\;\;\;\;\;\;\;\;\;\;\;\;\;\;\;\;\;
\;\;\;\;\;\;\;\;\;\;\;\;\;\;\;\;\;\;\;\;\;\;\;\;\;\;\;
\;\;\;\;\;\;\;\;\;\;\;\;+\lambda_1 \cos p s +\lambda_2 \sin p s=0.
\end{eqnarray}
These equations are  mutually compatible if and {\it only if} [20]
\begin{equation}
B^{2}=(A^{\prime}-A)C
\end{equation}
and the Lagrange multiplier $\bf{\lambda}$ is set to zero.
 Under  condition (9) we can get the 
sole solution of $\psi$ to be [17]
\begin{equation}
\psi^{\prime}=\omega_0-{\frac{2\pi B}{C L}}\sin\psi.
\end{equation}
with $p$ in Eqs.~(6-8) being replaced by $2\pi/L$.
Thus we know that the necessary and sufficient condition for an undistorted
closed DNA to form a flat circle is Eq.~(9). Note that
(9) is just also the compatibility condition
 between the  MS model and the WLC model.  So for the MS model to be appropriate,
in addition to the assumption that
unstressed linear DNA configurations have spatial twisting frequency $\omega_0$,
another fundamental assumption is needed. This assumption 
is that  unstressed closed DNA rings form flat circles, under 
which (9) will certainly be hold. 
Stable DNA circles have been observed in various experiments [18-20], 
so this assumption
is reasonable.

Correspondingly, we  derive the final form  of the elastic energy after
inserting Eq.~(9) into Eq.~(2), 
\begin{equation}
\beta E_c=\int ds [{\frac{A}{2}}(\Omega_1^{2}+\Omega_2^{2})+{\frac{C}{2}}(
\Omega_3-\omega_0+{\frac{B}{C}}\Omega_1)^{2}]
\end{equation}
Eq.~(11) is the central result of this paper.
It is interesting to see that, compared with the previously
mentioned achiral model (1), the only new thing  of Eq.~(11) 
is that a new term $(B/C)\Omega_1$ is added into the twisting energy. 
 In what follows we will
discuss the possible influences of this new term to DNA twisting manners.

For short equilibrium DNA rings, from Eq.~(10) we know that their HRL is
\begin{equation}
h=\{({\frac{\omega_0}{2\pi}})^{2}-({\frac{B}{ C L}})^{2}\}^{-1/2},
\end{equation}
which are longer than $h_0=2\pi/\omega_0$,
 the value for undistorted linear chains.
Especially interesting of Eq.~(12) is that it predicts 
that  the shorter the chain, 
the longer its HRL.  Such a tendency in HRL has been observed
in various experiments [18, 19], and its real nature has been controversial. 
The present work suggests it to be induced by the twisting energy in (11).

The case of long chains is  more important and interesting, but much more
difficult to tackle. Here we should use statistical methods. Model (11)
shows that for a chain with twisting freedom its axial (${\bf t}$)  distribution
is just that of a worm-like chain, with $\rho({\bf t}, {\bf t_0}, s)
\propto \int_{\bf t_0}^{\bf t} \exp[-(A/2)\int_0^L {{\bf t}^{\prime}}^2 ds]$, 
and for each specific axial configuration,
 $\langle {\psi}^{\prime}\rangle_{twist}=
\omega_0-{\phi}^{\prime}(s)\cos\theta-(B/C)\kappa(s) \sin(\psi+\alpha(s))$,
where $\langle\cdots\rangle_{twist}$ means average respect to twisting distribution
and $\alpha(s)=\arctan({\theta}^{\prime}/{\phi}^{\prime}
\sin\theta)$. Taking into
account the fact that $\psi$ changes much faster than $\phi$ and $\theta$,
we can take neglect the term ${\phi}^{\prime}\cos\theta$ and 
take $\kappa(s)=({\phi^{\prime}}^2 \sin^2\theta+{\theta^{\prime}}^2)^{1/2}$ as
constant  while calculating $\psi(s)$. Thus,
$\langle\dot{\psi}\rangle_{twist}\simeq \omega_0-(B/C)\kappa_{s}\sin(\psi+const)$
and the instantaneous HRL at arc-length $s$ is 
$h(s)=(2\pi/\omega_0)(1-B^2\kappa(s)^2/C^2\omega_0^2)^{-1/2}$. Consequently
the mean HRL is
\begin{equation}
\bar{h}=\langle h(s)\rangle_{bend}=h_0 [1+{B^2\over {2 C^2 \omega_0^2}}\langle
\kappa(s)^2\rangle_{bend}]
\end{equation}
where $\langle\cdots\rangle_{bend}$ means average with respect to the WLC chain.
We see from Eq.~(13)  that $\bar{h} >h_0$ whenever $B\neq 0$, in qualitative
agreement with experiment of Wang [15]. However, $\langle\kappa(s)^2\rangle_{bend}$
is difficult to calculate for a WLC chain, due to the fact that
$|{\bf t}|^{2}=1$. Here, we have to adapt a self-consistent field method to
convert this local constraint to a global one such that $\int_0^L {\bf t}^2 ds =L$
and determine the corresponding Lagrange multiplier self-consistently [21].
The self-consistent field internal energy is $
\beta E^{\prime}=\int_{0}^L (A/2) {{\bf t}^{\prime}}^2+\gamma {\bf t}^2+
{\bf \lambda}\cdot {\bf t} ]ds$ where ${\bf \lambda}$ is  for possible
end constraints. Detailed calculation shows that
\begin{equation}
\langle\kappa(s)^2\rangle_{bend}={9\over 4 A^2}
\end{equation}
which is independent of the value of ${\bf\lambda}$, i.~e., independent of
whether the chain is open (${\bf \lambda}=0$) or closed (${\bf \lambda}\neq 0$);
and it is independent of chain length as it should be. Thus Eq.~(13) predicts
that solution DNA, whether linear or circular, has the same mean HRL  longer than
that of rectilinear  DNA.
 
For very long closed DNA chains (about several ten thousand base pairs), gel
electrophoresis experiments did reveal  a significant increase in  mean 
HRL, with
$h=10.4$ bp [15], however, the HRL for rectilinear DNAs is only $10$  bp  [16].
As mentioned before, for a long time the real reason for this phenomenon is
not clear. Our present theory can give a natural and  reasonable explanation,
proposing that this discrepancy is induced by the chiral twisting energy
in model (11). To be more quantitative, we insert the experimental
values into Eq.~(13) and estimate that $B/C\simeq 17.8$.  This relatively
large value indicates
that $A^{\prime}$ is much higher than $A$ in Eq.~(2). 
We wish future experiments will check the results of our present theoretical
work.

As mentioned before, 
another interesting phenomenon observed in experiments is that in prokaryotic
cells and in some yeast cells many circular {\it plasmid} DNA molecules of several
thousand bp are found with linking numbers typically about $5$ per 
cent less
than that of the relaxed ring [19]. In fact, almost
all naturally occurred DNAs are negatively supercoiled, i.~e.,  with
deficient linking numbers. 
Why should natural DNA rings prefer a deficit rather than an equilibrium or
  an excess in linking
number? The real nature is still unknown, 
maybe this kind of bias is also related  with the chiral characteristics of 
DNAs described in model (11).  Investigations based on model (11) with topological
constraint in now in progress by the present authors. 

In summary, we have proposed a revised Marko-Siggia chiral elastic model 
for DNA molecules and discussed its predictions on DNA double helix mean
helical repeat length. The theoretical results show that for short DNA rings,
their mean HRL increases with the decreasing of chain length; while for
very long chains, whether open or closed, their mean HRL is independent of 
chain length and is longer the value for rectilinear DNAs. These results
are in good agreement with experiments. 
We expect future experiments can check our theoretical results.

This work is supported by the National Natural Science Foundation of China. 
We would like to thank Dr. Liu Quanhui, Dr. Yan Jie and Zhao Wei for
helpful communications.


\begin{references}
\bibitem{17}  The detailed derivation will appear in a full paper.
\end{references}
\end{document}